\documentclass[runningheads]{llncs}
\usepackage{graphicx}
\usepackage{program}
\usepackage{url}
\usepackage[T1]{fontenc}
\usepackage[ansinew]{inputenc}
\usepackage[caption=false]{subfig}
\begin{document}
\title{Arbitrary Scale Super-Resolution for Brain MRI Images\thanks{Supported by University of Cambridge.}}
%
%
\author{Chuan Tan \and
Jin Zhu \and
Pietro Lio'}
\authorrunning{Tan et al.}
%
\institute{University of Cambridge, Department of Computer Science and Technology, Cambridge CB3 0FD, United Kingdom \\
\email{\{ct538, jz426, pl219\}@cam.ac.uk}\\
}
\maketitle              
\begin{abstract}
Recent attempts at Super-Resolution for medical images used deep learning techniques such as Generative Adversarial Networks (GANs) to achieve perceptually realistic single image Super-Resolution. Yet, they are constrained by their inability to generalise to different scale factors. This involves high storage and energy costs as every integer scale factor involves a separate neural network. A recent paper has proposed a novel meta-learning technique that uses a Weight Prediction Network to enable Super-Resolution on arbitrary scale factors using only a single neural network. In this paper, we propose a new network that combines that technique with SRGAN, a state-of-the-art GAN-based architecture, to achieve arbitrary scale, high fidelity Super-Resolution for medical images. By using this network to perform arbitrary scale magnifications on images from the Multimodal Brain Tumor Segmentation Challenge (BraTS) dataset, we demonstrate that it is able to outperform traditional interpolation methods by up to 20$\%$ on SSIM scores whilst retaining generalisability on brain MRI images. We show that performance across scales is not compromised, and that it is able to achieve competitive results with other state-of-the-art methods such as EDSR whilst being fifty times smaller than them. Combining efficiency, performance, and generalisability, this can hopefully become a new foundation for tackling Super-Resolution on medical images.

\keywords{Super-Resolution  \and Medical Image Analysis \and Meta-Learning \and Image Processing}
\end{abstract}
\section{Introduction}
In this paper, we seek to apply elements of Meta-Learning to Generative Adversarial Networks (GANs) to tackle Super-Resolution in medical images, specifically brain MRI images. We are the first to apply such a scale-free Super-Resolution technique on these images. Super-Resolution is the task of increasing the resolution of images. It helps radiological centres located in rural areas which do not have high-fidelity instruments achieve comparable diagnostic results as their advanced counterparts in the city. The importance of Super-Resolution is growing as cross modality analysis requires combining different types of information (such as PET and NMR scans) of varying resolution.

Traditionally, Super-Resolution has been done using interpolation, such as bicubic interpolation, but recent attempts have involved the use of deep learning methods to extract high level information from data, which can be used to supply additional information to increase the resolution of the image. Current deep learning techniques for medical images rely heavily on Generative Adversarial Networks (GANs) for they are able to generate realistic and sharper images by using a different loss function that yields high perceptual quality \cite{GANsinmedicalimaging}. For example, mDCSRN \cite{GANsinmedicalimaging}, Lesion-focussed GAN \cite{GANsforSR_1}, ESRGAN \cite{GANsforSR_2}, are all GAN-based solutions tackling Super-Resolution for medical images. Most of them are based off SRGAN - a GAN-based network tackling Super-Resolution developed by Ledig et al. \cite{DBLP:journals/corr/LedigTHCATTWS16}. We therefore use SRGAN as a foundation to which we apply our modification.

Despite their better performance, almost all the networks relied on the sub-pixel convolution layer introduced by Shi et al. \cite{DBLP:journals/corr/ShiCHTABRW16}, which tied a particular upscaling factor to a particular network architecture. This incurs high storage and energy costs for medical professionals who may wish to conduct Super-Resolution on different scaling factors. In this paper, we combine the Meta-Upscale Module introduced by Hu et al. \cite{DBLP:journals/corr/abs-1903-00875} with SRGAN to create a novel network lovingly termed Meta-SRGAN. This breaks the constraint imposed by Shi et al's layer and is capable of tackling Super-Resolution on any scale, even non-integer ones, and hence can reduce storage, energy costs and lay the foundations for real-time Super-Resolution. We first show that Meta-SRGAN outperforms the baseline of bicubic interpolation on the BraTS dataset \cite{BraTSCite1,BraTSCite2,BraTSCite3}, and also show that Meta-SRGAN is capable of performing similarly to SRGAN, but yet is able to super-resolve images of arbitrary scales. We also compare the memory footprint and show that Meta-SRGAN is 
$\approx$ 98$ \% $ smaller than EDSR, a state-of-the-art Super-Resolution technique, but yet is able to achieve similar performance.

The rest of the paper is organised as follows. Background introduces Super-Resolution formally and illustrates how Hu et al. reframes the problem. We then introduce the architecture of Meta-SRGAN in Methods. Experiments elaborates on dataset preparation and training details. Finally, we show the outcomes of the experiments in Results.

\subsection{Background}
\subsubsection{Super-Resolution}
\begin{figure}
	\includegraphics[width=\linewidth]{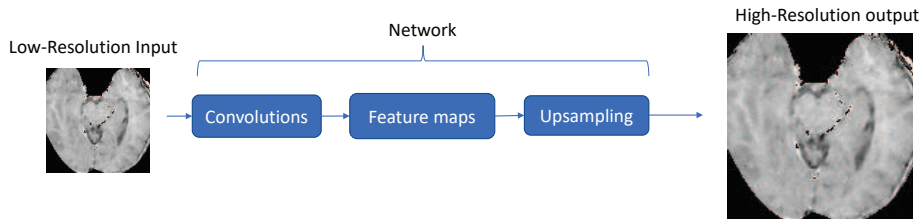}
	\caption{How a Super-Resolution task is often carried out. A low-resolution input image undergoes some convolutions to generate feature maps. Feature maps are some sort of representation that captures the features of the low-resolution image. Depending on the upscaling factor (it must be a multiple of $2^n$), there would be some number $n$ of upscaling modules applied to the representation. This upscaling unfolds the image into the high-resolution output. }
	\label{fig:SRTask}
\end{figure}

Deep learning-based Super-Resolution techniques are generally supervised learning algorithms. They analyse relationships between the low-resolution (LR) image and its corresponding high-resolution (HR) image, and use this relationship to obtain the super-resolved (SR) image. This SR image is then evaluated against the HR image to see how well the algorithm is performing. Any algorithm that is capable of extracting these relationships can be used to do Super-Resolution. Dong et al \cite{DBLP:journals/corr/DongLHT15} showed that any such algorithms can be thought of as convolutional neural networks (CNNs). Figure \ref{fig:SRTask} shows a typical high level recipe for a Super-Resolution task.

\subsubsection{Meta-Upscale Module}
\begin{figure}
	\includegraphics[width=\linewidth]{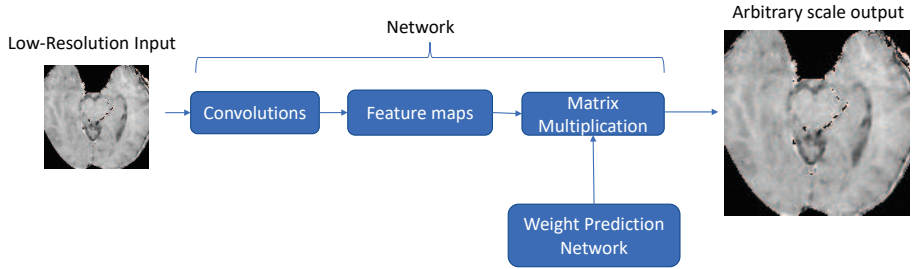}
	\caption{The framework for tackling Super-Resolution tasks, as proposed by Hu et al. There is no constraint on the upscaling factor. Note that the Upsampling module has been replaced with a matrix multiplication and a Weight Prediction Network. The matrix multiplication and Weight Prediction Network are collectively called the Meta-Upscale Module.}
	\label{fig:ArbitraryScaleSRTask}
\end{figure}

The Upsampling module in the network is often the efficient sub-pixel convolution layer proposed by Shi et al. \cite{DBLP:journals/corr/ShiCHTABRW16}. Instead of explicitly enlarging feature maps, it expands the channels of the output features to store the extra points to increase resolution. It then rearranges these points to obtain the super-resolved output image. Almost every network tackling Super-Resolution uses this Upsampling module.  An example is EDSR \cite{DBLP:journals/corr/LimSKNL17}, a convolutional neural network tackling Super-Resolution. Hu et al. then proposed replacing the Upsampling module with a Meta-Upscale Module \cite{DBLP:journals/corr/abs-1903-00875}. This Meta-Upscale Module consists of a Weight Prediction Network and is an example of Meta-Learning as it generalises a network to tackle more than one upscaling factor. Essentially, given a few training examples (small amount of images from some scales), we train a network to tackle arbitrary scale factors. The Weight Prediction Network is able to predict weights for different upscaling factors. This extra layer of abstraction enables the underlying neural network to generalise across tasks. The framework proposed by Hu et al. for enabling arbitrary scale Super-Resolution is illustrated in Figure \ref{fig:ArbitraryScaleSRTask}.

Hu et al. reframes Super-Resolution as a matrix multiplication problem:\begin{equation} Y = Wx + b \end{equation} where $Y$ is the super-resolved image, $W$ is some matrix of weights we learn, $x$ is the input low-resolution image, and $b$ is some constant. This may seem like an oversimplification and in many ways it is (for example it doesn't take account how pixels close to each other tend to exert more influence on each other), but it gives us a rather intuitive understanding for why a separate weight prediction network works for arbitrary scale Super-Resolution. Consider $x$ as a $ 4 \times 1 $ vector, and $W$ as a $ 8 \times 4 $ matrix. The output $Y$ is therefore a $ 8 \times 1 $ vector. In some sense we have doubled the size of $x$. We can think of W as providing an upscaling of $2$. The idea proposed by Hu et al. is to predict W, for every upscaling factor. This is done by passing in the input dimensions to an external function to create the shape of W, before using a weight prediction network to predict the values of $W$, as in the equation below:
\begin{equation} W = f_{\theta}(x)\end{equation}
where $W$ is the matrix of weights, $f_{\theta}$ is the weight prediction network with parameters $\theta$, and $x$ is the input low-resolution image.

\begin{figure}[h]
	\centering
	\includegraphics[width=.8\linewidth]{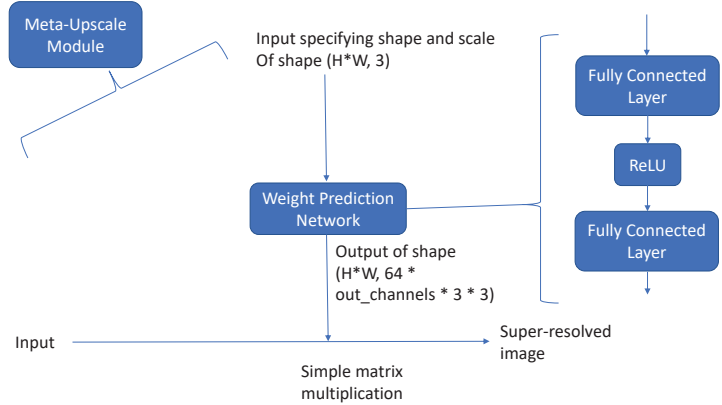}
	\caption{Architecture of Meta-Upscale Module. $H$ and $W$ refers to the height and width of the \textit{high-resolution} image, and $out\_channels$ refers to the number of channels in the original image (typically $3$ for colour, $1$ for grayscale). $64$ refers to the number of channels in the input features fed into the Meta-Upscale Module. $3$ refers to the kernel size used in the last convolution layer in the network producing the input features fed into the Meta-Upscale Module.  }
	\label{fig:metardn_architecture_3}
\end{figure}

The architecture of the Meta-Upscale Module is shown in Figure \ref{fig:metardn_architecture_3}, and includes a weight prediction network that outputs weights which are then matrix multiplied to the input features to obtain the super-resolved image. 
The Weight Prediction Network consists of three layers, and is trained alongside the main network. Of significant importance is the shape of the input into the Weight Prediction Network. It must contain information relating to the shape of the high-resolution image, as that determines the size of the weight matrix that is multiplied with the input features fed into the Module. Additional details can be found in \cite{DBLP:journals/corr/abs-1903-00875}.

\section{Methods}
\subsection{Baselines}
We used Bicubic Interpolation as a baseline, and implemented EDSR as a state-of-the-art technique that we compare Meta-SRGAN to. We also wanted to investigate how well using the Meta-Upscale Module on a GAN will affect its performance, so we implemented SRGAN too. EDSR and SRGAN were trained on x2 upsampling tasks, and were trained on DIV2K for 160k updates before transferring their learning onto the BraTS dataset for an additional 160k updates. EDSR was trained with 256 number of features, 32 residual blocks, and a residual scaling of 0.1 (see \cite{DBLP:journals/corr/LimSKNL17} for more information). EDSR was trained using L1 Loss, whilst SRGAN was trained using the same combination of losses used to train Meta-SRGAN.

\subsection{Meta-SRGAN}
To reap the benefits of both GANs and meta-learning, we combine Ledig et al.'s SRGAN with Hu et al.'s Meta-Upscale Module to obtain a new architecture called Meta-SRGAN. The nature of a GAN enables the generator to generate realistic images. The ability to predict weights for different upscaling factors enables the network to tackle arbitrary scales. Combined, they result in a network that is both hugely powerful and highly generalisable.

\begin{figure}[h]
	\centering
	\subfloat{%
		\label{subfig:metasrgan_generator}
		\includegraphics[clip,width=0.9\columnwidth]{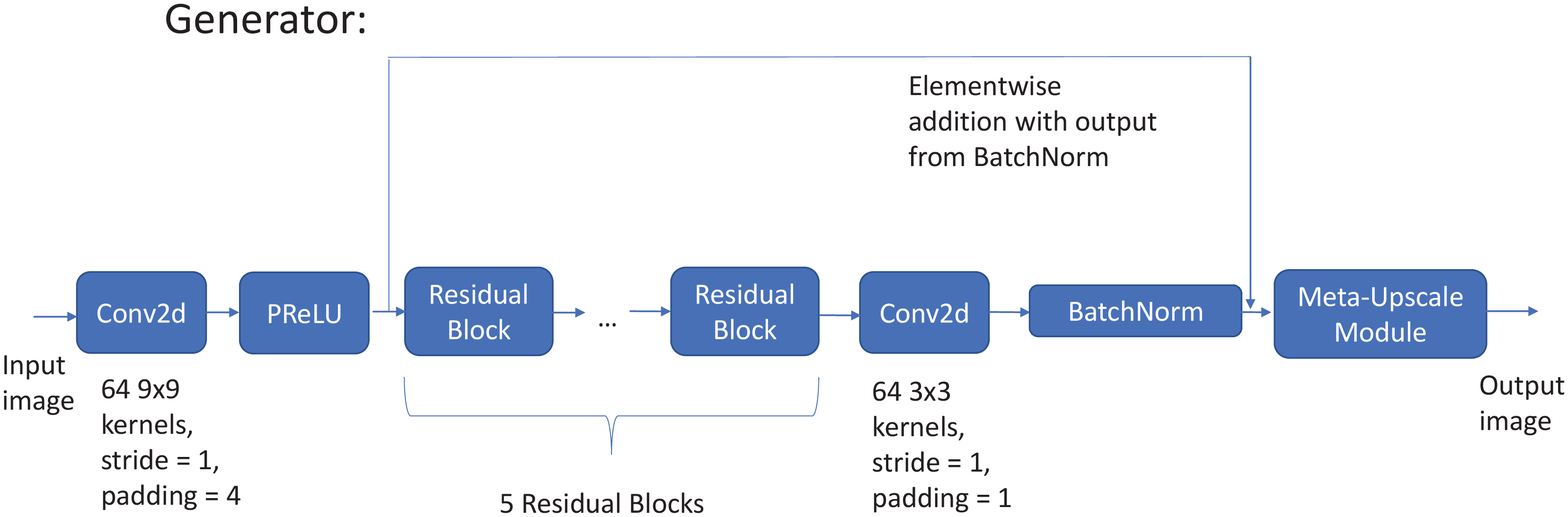}%
	}
	
	\subfloat{%
		\label{subfig:metasrgan_residualblock}
		\includegraphics[clip,width=0.7\columnwidth]{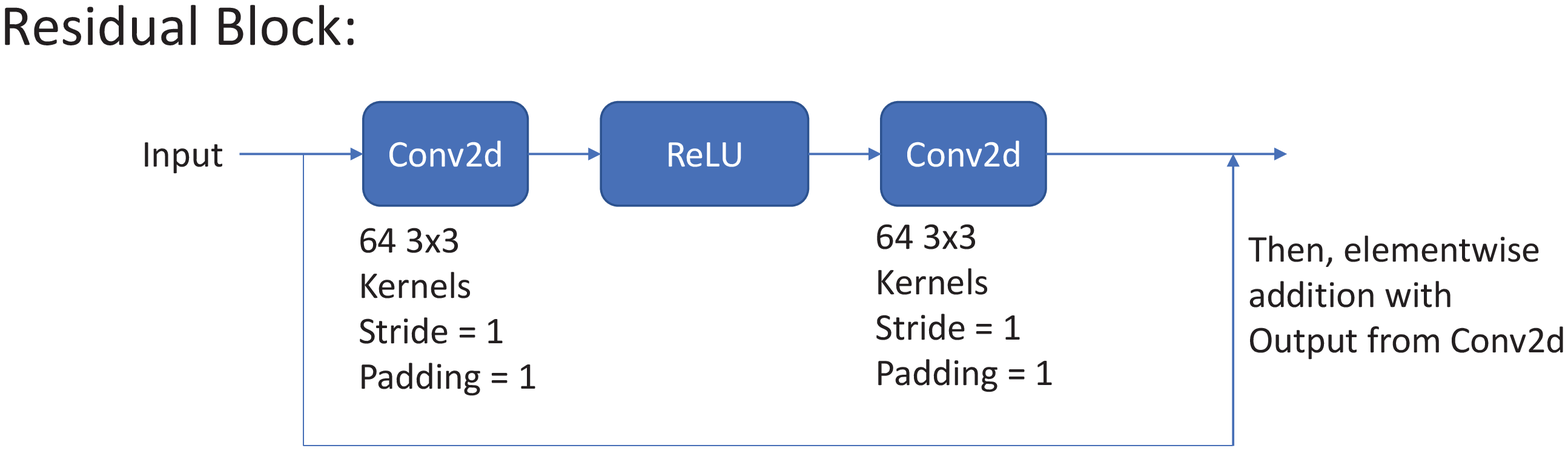}%
	}
	\caption{(Top) Architecture of Generator. Generator consists of Convolutions, Residual Blocks, Batch Normalisation, and the Meta-Upscale Module. \textit{PReLU} refers to the Parametric Leaky ReLU activation function. (Bottom) Architecture of Residual Block used in Generator. \textit{ReLU} refers to the ReLU activation layer. The number 64 refers to the number of kernels, and can be modified like any hyperparameter. }
	\label{fig:metasrgan_architecture_1}
\end{figure}

\begin{figure}[h]
	\centering
	\includegraphics[width=.9\linewidth]{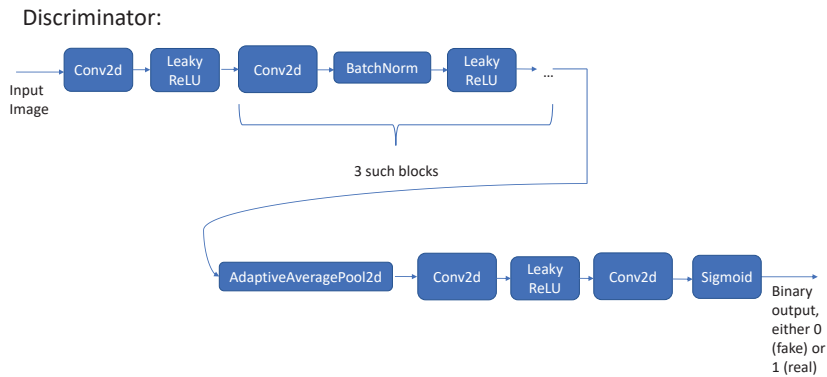}
	\caption{Architecture of Discriminator. \textit{Leaky ReLU} refers to the Leaky ReLU activation function. \textit{AdaptiveAveragePool2d} refers to an average pooling layer, whilst \textit{Sigmoid} refers to the sigmoid activation function that outputs a binary value.}  
	\label{fig:metasrgan_architecture_2}
\end{figure}

There are two networks being trained in a Generative Adversarial Network. There is the \textit{generator} which generates an image (in this case it super-resolves an image) which is then fed into the \textit{discriminator} which discriminates between a real and a generated image (in this case it determines whether the image presented is the ground truth high-resolution, or not). This feedback from the discriminator is then used to further train the generator. Both generator and discriminator are playing a game to outbid each other. The generator tries to improve its ability to generate images that can fool the discriminator whilst the discriminator tries to improve its ability to discern real images from generated images. This \textit{adversarial learning} enables us to generate more realistic and detailed images. The architecture of the generator and discriminator are shown in figures \ref{fig:metasrgan_architecture_1} and \ref{fig:metasrgan_architecture_2} respectively.

\subsubsection{Generator}
The generator is used to generate super-resolved images, which the discriminator will take in and output a classification that says whether it is a generated image or a real image.
Residual Blocks are used in the generator to stabilise training by incorporating a feedback loop back onto the input. Parametric ReLU was used as it was empirically found to stabilise training \cite{DBLP:journals/corr/LedigTHCATTWS16}. 
\subsubsection{Discriminator}
The discriminator consists of several Leaky ReLU layers and an average pooling layer. These was empirically found to stabilise training \cite{DBLP:journals/corr/LedigTHCATTWS16}. A sigmoid layer was used to facilitate binary classification.
\subsubsection{L1 Loss} 
L1 Loss is the typical loss function used by other networks that calculates pixel-wise errors. We incorporate it to train the generator in Meta-SRGAN. We denote an input image by $x$, and a neural network and its parameters by $f_{\theta}$. Letting $\hat{y} = f_{\theta}(x)$, $i$ be just an index, $x_i$ be the low-resolution input image and $y_i$ be the high-resolution target image, $C \times H \times W$ be the dimensions of an image, and $\vert \vert\dots\vert \vert$ be the Frobenius norm, L1 Loss is defined as
\begin{equation}
\ell_{l1}(\hat{y}, y) = \frac{1}{CHW} \sum_i \(\vert \vert\) \hat{y}_i - y_i \(\vert \vert\)
\end{equation}
L1 Loss minimises the mean absolute error between pixels and helps Meta-SRGAN match its output as closely as possible to the high-resolution target. 

\subsubsection{Adversarial Loss}
The generator tries to minimise the following function whilst the discriminator tries to maximise it. $D(x)$ is the discriminator's output that the high-resolution image is high-resolution, $G(z)$ is the super-resolved image produced by the generator, and $D(G(z))$ is the discriminator's output that the super-resolved image is high-resolution.
\begin{equation}
E_x[log(D(x))] + E_z[log(1-D(G(z)))]
\end{equation}
The above function is termed Adversarial Loss, and it allows Meta-SRGAN to achieve realistic super-resolved images.
 
\subsubsection{Perceptual Loss}
Meta-SRGAN also incorporates Perceptual Loss, a type of loss introduced by Johnson et al. to improve performance on Super-Resolution tasks \cite{JohnsonAL16}. The idea was to encourage the generated image to have similar feature representations as computed by a separate network. We use a 19-layer VGG network \cite{simonyan2014deep} pretrained on the ImageNet dataset\cite{RussakovskyDSKSMHKKBBF14}, and we extract the features before the last layer of the network. Letting $\hat{y}$ and $y$ be the super-resolved and high-resolution images respectively, $\phi$ be the network that extracts their feature representations, and assuming that the feature representations are of shape $C \times H \times W$, we define the Perceptual Loss as the Euclidean distance between the feature representations:
\begin{equation}
\ell_{p}(\hat{y} , y) = \frac{1}{CHW} \vert\vert \phi(\hat{y}) - \phi(y) \vert\vert^2
\end{equation}

In the following experiments we combined the loss functions as
\begin{equation}
\ell_{total}(\hat{y} , y) = \ell_{l1}(\hat{y}, y) + 0.001 \ell_{adversarial}(\hat{y}, y) + 0.006 \ell_{p}(\hat{y}, y) 
\end{equation}
and use it to train Meta-SRGAN.

\section{Experiments}
We performed experiments on the Multimodal Brain Tumor Segmentation (BraTS) dataset \cite{BraTSCite1,BraTSCite2,BraTSCite3}. We first trained Meta-SRGAN on the DIV2K dataset \cite{Agustsson_2017_CVPR_Workshops} - a dataset curated for Super-Resolution tasks, and then transferred that learning onto the BraTS dataset.
\subsection{Datasets and Data Preparation}
To demonstrate how well Meta-SRGAN performs on medical images, we trained and tested Meta-SRGAN on the Multimodal Brain Tumor Segmentation (BraTS) dataset \cite{BraTSCite1,BraTSCite2,BraTSCite3}. This dataset contains several versions, including t1, t1ce, and t2 brain MRI scans, and serves as a good proxy for medical images. We picked the 2D t1ce version as it was two dimensional and was simple to work with. There were $15140$ training images and $3784$ validation images. Training images were normalised using a mean and standard deviation of $0.370$ and $0.117$ respectively before they were passed to the network to be trained on. These values were calculated using pixel values from the training images, after the maximal informational crop was applied.
\begin{figure}[h]
	\centering
	\includegraphics[width=.9\linewidth]{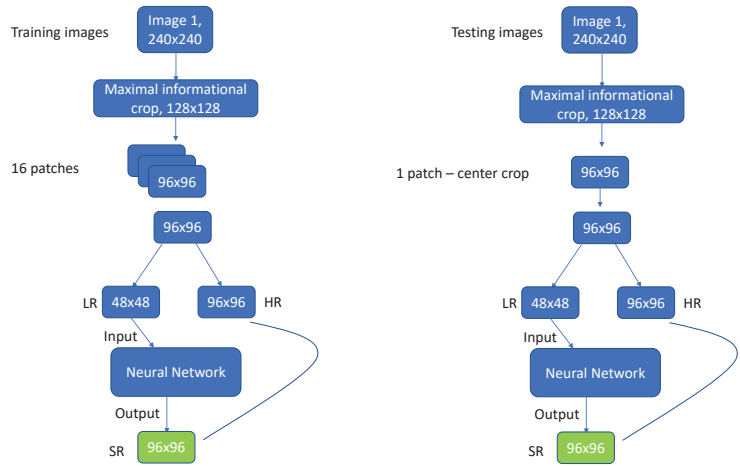}
	\caption{Workflow for handling the BraTS dataset, for a x2 upsampling task.}
	\label{fig:BraTS_workflow}
\end{figure} 
\noindent
The BraTS dataset is tricky to deal with because every image has sparse information. Only about $50\%$ of each image contained useful information (i.e. the brain), the rest was black. To tackle this problem, we cropped the image with the most pixel information. This corresponded to crops of the brain. Then, from this crop, we then randomly sampled $16$ $96$ by $96$ patches to provide sufficient coverage per image. For an upscaling factor of 2, the low-resolution input image (LR) was of size $48$ by $48$ whilst the high-resolution target image (HR) and super-resolved output image (SR) were of size $96$ by $96$. This workflow is summarised in Figure \ref{fig:BraTS_workflow}.
\subsection{Training Parameters and Experiment Settings}
We first trained Meta-SRGAN on the DIV2K dataset for 360k updates, with a mini-batch size of 8. We then did transfer learning and trained Meta-SRGAN on the BraTS dataset for an additional 360k updates. We trained Meta-SRGAN on a range of scales from 1.1 to 4.0 (with 0.1 increments). Higher upscaling factors were not chosen as that would have made the input image too small to be useful. Each minibatch of images was associated with a scale, and each scale was chosen uniformly at random from 1.1 to 4.0. We optimised the network using a combination of L1 Loss, Adversarial Loss, and Perceptual Loss as described in the previous section. We used an Adam optimiser with a learning rate of $1 * 10^{-4}$ on each of the generator and discriminator, and decreased the learning rate by $20\%$ every 60k updates. Pixel values were clamped to between 0 and 255 to give the generator an edge. The network was trained on a NVIDIA TITAN X GPU and took three days. We calculated the PSNR and SSIM scores only on the luminance channel of the maximal information crops.

\section{Results}
\begin{figure}[h!]
	\centering
	\includegraphics[width=\linewidth]{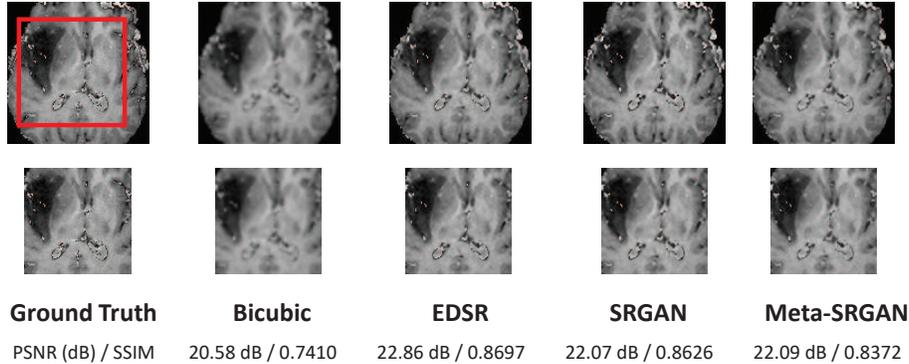}
	\caption{Sample images from Bicubic Interpolation, EDSR, SRGAN and Meta-SRGAN. This is from a x2 upsampling task. }
	\label{fig:comparison_x2_sample_images}
\end{figure}

The results of our experiments are tabulated in Table \ref{table_metasrgan_brats}. We see that Meta-SRGAN clearly outperformed the baseline of Bicubic Interpolation, which suggests that the inclusion of the Meta-Upscale Module did not hinder the network's performance. This is further reinforced by the fact that Meta-SRGAN was able to achieve similar performance on the x2 upsampling task as SRGAN, the architecture it was based on. It also had the added benefit of a smaller number of parameters compared to SRGAN. Visual quality is indicated by a proxy metric, Structural Similarity Index Metric (SSIM), which measures the perceived image similarity between two images. The high SSIM scores achieved by the GANs bodes well for brain MRI imaging which demands accurate images. We also see that Meta-SRGAN has the lowest number of parameters, which means its memory footprint is really small ($\approx$ 98$ \% $ smaller than EDSR), which suggests that it can be deployed easily.
\begin{table}[]
	\centering
	\caption{A comparison of PSNR and SSIM scores between Bicubic Interpolation, SRGAN and Meta-SRGAN on the BraTS dataset. Best results are \textbf{bolded}.}
\resizebox{\textwidth}{!}{%
	\begin{tabular}{|c|c|c|c|l|c|l|c|l|}
		\hline
		\multicolumn{9}{|c|}{x2 Upsampling Task, BraTS Testing Set}                                                                                            \\ \hline
		& \multicolumn{2}{c|}{Bicubic Interpolation} & \multicolumn{2}{c|}{EDSR}  & \multicolumn{2}{c|}{SRGAN}  & \multicolumn{2}{c|}{Meta-SRGAN} \\ \hline
		& PSNR (dB)             & SSIM               & PSNR (dB)        & SSIM    & PSNR (dB) & SSIM            & PSNR (dB)        & SSIM         \\ \hline
		mean         & 17.98                 & 0.6091             & \textbf{18.93}   & 0.6885  & 18.42     & \textbf{0.7295} & 18.72            & 0.7279       \\ \hline
		std          & 5.674                 & 0.1708             & 4.2909           & 0.1814  & 5.4975    & 0.1308          & 5.338            & 0.1422       \\ \hline
		\#parameters & \multicolumn{2}{c|}{-}                     & \multicolumn{2}{c|}{40.7M} & \multicolumn{2}{c|}{0.565M} & \multicolumn{2}{c|}{0.561M}     \\ \hline
	\end{tabular}%
}
\label{table_metasrgan_brats}
\end{table}
Sampled images are shown in Figure \ref{fig:comparison_x2_sample_images}. The higher SSIM scores of SRGAN and Meta-SRGAN than the baseline of Bicubic Interpolation highlights the fact that they generate perceptually better images than the baseline. The very fact that Meta-SRGAN is even able to have comparable scores to EDSR despite being trained on a whole range of scales, and with considerably less parameters, attests to not just the power of adversarial training, but also the robustness of the Meta-Upscale Module.

\begin{figure}[h!]
	\centering
	\includegraphics[width=\linewidth]{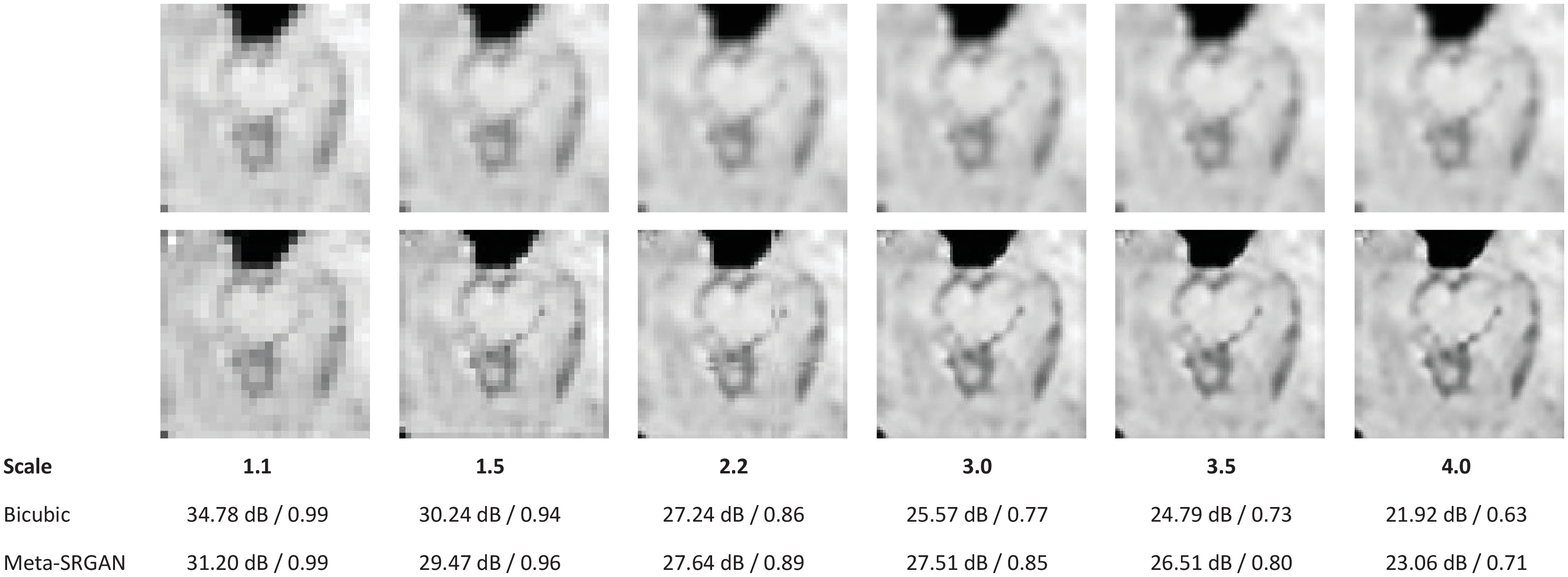}
	\caption{(Top) Image going through various upsampling factors using Bicubic Interpolation. (Bottom) Image going through various upsampling factors using Meta-SRGAN.  PSNR and SSIM scores are calculated against a target image downscaled by a factor of 4.}
	\label{fig:metasrgan_sequence_upsampling_factors_brats}
\end{figure}

Meta-SRGAN was also able to produce a sequence of images corresponding to different upscaling factors. This is shown in Figure \ref{fig:metasrgan_sequence_upsampling_factors_brats}. The relatively consistent PSNR and SSIM scores indicate that performance across different scales are not compromised. All these results suggests that we have a memory efficient network capable of generating super-resolved brain MRI images of high visual quality, and can be used to generate images of any scale.

With better visual quality than the baseline, lower memory footprint, and ability to tackle any upscaling factor with negligible loss in performance, Meta-SRGAN clearly can deliver comparable performance to state-of-the-art Super-Resolution techniques.

\section{Conclusions}
The implications of the results on the BraTS dataset are three-fold. The low memory footprint of Meta-SRGAN enables it to be wrapped as a helper tool to aid medical tasks such as MRI and endoscope videography. The ability to super-resolve images on any arbitrary scale allows one to easily extend this to an application of real-time zooming and enhancing of an image, which will be useful in medical screening and surgical monitoring. The fact that its performance is not hindered across multiple upscaling factors affords it the flexibility to be used in other works such segmentation, de-noising, and registration. Despite only testing it on brain MRI images, Meta-SRGAN can easily be extended to datasets of other modalities such as Ultrasound and CT scans. Its low memory footprint, high visual quality, and generalisability suggests that Meta-SRGAN can become a new foundation on which other architectures can be built to enhance the performance for medical images. 

We have built a network that combines the ability to generate images of high visual quality with the ability to tackle arbitrary scales, and are the first to show that this does not compromise performance or memory footprint for brain MRI images. This means that unlike other state-of-the-art methods, our method works on arbitrary scales which means that only a single network is required to perform Super-Resolution on any upscaling factors. In future work we hope to enhance the performance of Meta-SRGAN and apply it to other modalities.

\section{Acknowledgements}
Jin Zhu’s PhD research is funded by China Scholarship Council \\(grant No.201708060173), whilst Pietro Lio' is supported by the GO-DS21 EU grant proposal.
\bibliographystyle{splncs04}
\bibliography{refs.bib}
\end{document}